\documentclass{appolb}
\usepackage{graphicx}

\begin{document}
\title{The AFP and PPS projects%
}
\author{Christophe Royon
\address{IRFU-SPP, CEA Saclay, F91191 Gif-sur-Yvette cedex}
\\
{Nicolo Cartiglia
}
\address{INFN Torino,  Via Pietro Giuria 1, Torino, Italy}
}
\maketitle
\begin{abstract}
We present the project to install new forward proton detectors in the CMS and
ATLAS experiments called PPS and AFP respectively.
\end{abstract}

\section{The AFP and CT-PPS projects}

\begin{figure}[htb]
\centerline{%
\includegraphics[width=10.5cm]{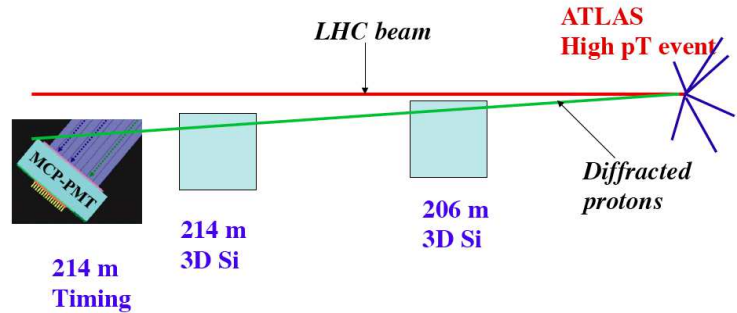}}
\caption{Scheme of the AFP proton detector in ATLAS. The same detector is
implemented on the other side of ATLAS.}
\label{Fig1}
\end{figure}

The Large Hadron Collider (LHC) will collide protons with a center-of-mass
energy of 13 TeV starting in 2015. Several improvements are made to the ATLAS and CMS 
detectors to exploit this new energy regime; this short report describes 
the project to install the ATLAS Forward Proton (AFP) detector at 206 and 214 
meters on both sides of the ATLAS experiment~\cite{loi} (see Fig.~\ref{Fig1}) 
and the similar project by the TOTEM and CMS collaborations, the so called CT-PPS, 
to be installed on both sides of the CMS detector.  In this article, we will 
concentrate on the main characteristics of the AFP and CT-PPS detectors, 
while their physics reach is described elsewhere~\cite{physics,physicsb}..

Each arm of the AFP detector will consist of two sections: AFP1 at 206 meters, 
and AFP2, at 220 meters. In AFP1, a tracking station composed by 6 layers of 
Silicon detectors will be deployed. The second section, AFP2, will contain a 
second identical tracking station and a timing detector. In addition, a similar 
structure could be installed at about 420 m from the ATLAS interaction point. 
The aim of the combined two arms of this setup is to tag the protons emerging intact 
from the pp interactions,  allowing ATLAS to exploit the program of diffractive 
and photon induced processes described in the previous sections. Likewise, 
the CT-PPS of CMS will also use the same combination of tracking and timing 
detectors, with the far station using specially designed cylindrical roman 
pots to house the timing detectors. 

\begin{figure}
\centering
\includegraphics[width=3.5in]{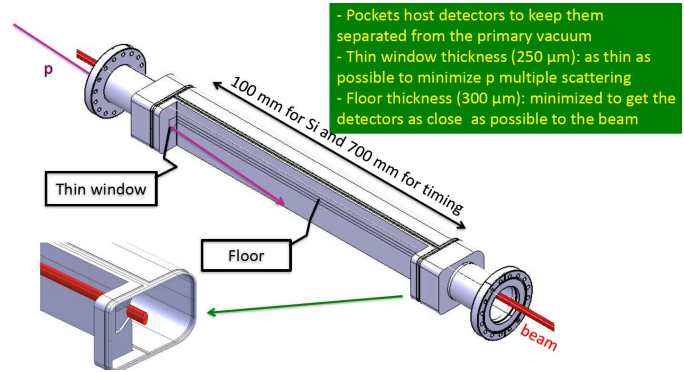}
\caption{Scheme of the movable beam pipe.}\label{Figa}
\end{figure}

\section{Movable beam pipes and roman pots}
In order to house the detectors needed by the AFP and CT-PPS projects,  two 
different types of modification of the beam-pipe are currently considered:  
(i) roman pots and (ii) movable beam pipes). Roman pots have been used
already in many experiments at the SPS, HERA, Tevatron and LHC colliders (in
the TOTEM and ATLAS-ALFA experiments). The roman pots, in their basic design, 
are pockets
where the detectors can be hosted. These pockets are pushed inside the beam 
pipe to a position closer to the beam line once
stable beam has been declared (a typical motion is of the order of a few cm). 
To minimize multiple scattering, protons will enter the roman pots via  
a thin window located at the bottom of the pot, on the side facing the beams.  
Different types of roman pots can host the tracking  and timing 
detectors: tracking detectors need less space than timing detectors, 
and therefore can be housed in smaller roman pots. 
. 
Conversely, in the movable beam pipe design, no pocket is pushed
closer to the beam, but the whole beam pipe moves closer to the beam.
The idea of movable Hamburg beam pipes is quite simple~\cite{HBP}: a section 
of the LHC beam pipe is replaced by a larger one, specially designed 
to have a cutout to host the detectors. When allowed by beam conditions, 
using specially designed bellows that allow for transverse motions, 
this part of the beam pipe is moved, by about 2.5 cm, 
so that the detectors located at its edge (called pocket)
are brought closer to the beam. 
In its design, the 
most challenging aspect is the minimization of the thickness of the portions 
called floor and window (see Fig.~\ref{Figa}). This is necessary as the 
floor might be rather long, of the order of 10-40 centimeters in the 
direction parallel to the motion of the particles: 
minimizing its depth of the floor ensures that the
detectors can be brought as close to the beam as possible allowing detecting 
protons scattered at very small angles. 
Two configurations exist for the movable beam pipes: the first one at 206 m from the ATLAS interaction
point hosts a Si detector 
(floor length of about 100 mm) and the second one (floor 
length of about 400 mm) the timing and the Si detectors. 

The AFP and CT-PPS detectors will use Roman Pots in their starting
configuration. In the meantime, the development of the Hamburg beam pipe is 
carried on together by both collaborations. 
However, it is clear that movable beam
pipes are needed at 420 m, if later upgrades include installation of 
forward detectors at that location. At 420 m,
not enough space is available and
new specially designed cryostats have been developed to host these movable 
beam pipes in the cold region.
The usage of roman pots at 420 m would require a costly cryogenic bypass to be 
installed to isolate
the part of the beam pipe where roman pots would be installed.

\section{3D Silicon detectors}
The purpose of the tracker system is to measure points along the trajectory of 
beam protons that are deflected at small angles as a result of collisions. 
The tracker, when combined with the LHC dipole and quadrupole magnets,
forms a powerful momentum spectrometer. Silicon tracker stations will be
installed in  Hamburg beam pipes or roman pots at $\pm$ 206 and $\pm$ 214 m from the 
ATLAS interaction point (and also at 420 m later if these additional detectors are appoved). 

The key requirements for the silicon  tracking system at 220~m are:
\begin{itemize}
\item Spatial resolution of $\sim$ 10 (30) $\mu$m per detector station in $x$ ($y$)
\item Angular resolution for a pair of detectors of a few $\mu$rad
\item High efficiency over the area of 20~mm $\times$ 20~mm corresponding to the distribution of
diffracted protons
\item Minimal dead space at the edge of the sensors towards the beam line, 
allowing measuring the scattered protons at
low angles
\item Sufficient radiation hardness in order to sustain the radiation at 
high luminosity
\item Capable of robust and reliable operation at high LHC luminosity 
\end{itemize}

The basic building unit of the AFP detection system is a module consisting of 
an assembly of a sensor array, on-sensor read-out chip(s), electrical services,
data acquisition and detector control system. The module will be
mounted on the mechanical support with embedded cooling and other necessary
services. The sensors are double sided 3D 50$\times$250 micron pixel detectors with slim-edge 
dicing built by FBK and CNM companies. The sensor efficiency has been measured to be close to 100\% 
over the full size in beam tests. A possible upgrade of this device will be to
use 3D edgeless Silicon
detectors built in a collaboration between SLAC, Manchester, Oslo, Bergen...

A new 
front-end chip FE-I4 has been developed for the Si detector by the Insertable B Layer (IBL)
collaboration in ATLAS~\cite{IBL}. The FE-I4 integrated circuit contains 
readout circuitry for 26 880 hybrid pixels arranged in 80~columns on 250~$\mu$m
pitch by 336 rows on 50 $\mu$m pitch, and covers an area of about 19 mm 
$\times$ 20 mm. It is designed in a 130 nm feature size bulk CMOS process. 
Sensors must be DC coupled to FE-I4 with negative charge collection. The FE-I4 
is very well suited to the AFP requirements: the granularity of cells provides 
a sufficient spatial resolution, the chip is radiation hard enough 
up to a dose of 3~MGy, 
and the size of the chip is sufficiently 
large that one module can be served by just  one chip. 

The dimensions of the individual cells in the FE-I4 chip are 50 $\mu$m $\times$
250 $\mu$m in the  $x$ and $y$ directions, respectively.
Therefore to achieve the required position resolution in the
$x$-direction of $\sim$ 10 $\mu$m, six layers with sensors are required
(this gives  50/$\sqrt{12}$/$\sqrt{5}\sim 7$ $\mu$m in $x$ and roughly 5 times 
worse in $y$). Offsetting planes alternately to the left and right by one half 
pixel will give a further reduction in resolution of at least 30\%. 
The AFP sensors are expected to be exposed to a dose of 30~kGy
per year at the full LHC luminosity of 10$^{34}$cm$^{-2}$s$^{-1}$.

The baseline CT-PPS tracking system is also based on 3D pixel sensors, 
produced either by FBK (Trento,  Italy) or CNM (Barcelona, Spain), 
which provide the best performance in terms of active  region and radiation 
hardness.

The chosen configuration for the tracking system consists of two detector
stations in each arm. Each station will contain one stack of silicon tracking
detectors. Each stack will consist of six planes, where each plane conatains a
1.6 $\times$ 2.4 cm$^2$ pixel sensor read out by six PSI46dig readout chips
ROCs~\cite{cms1}. Each ROC reads 52 $\times$ 80 pixels with dimensions 150
$\times$ 100 $\mu$m$^2$. The design of the front-end electronics and of the DAQ
is based on that developed for the Phse I upgrade of the CMS silicon pixel
detectors~\cite{cms2}.

\begin{figure}
\centering
\includegraphics[width=3.5in]{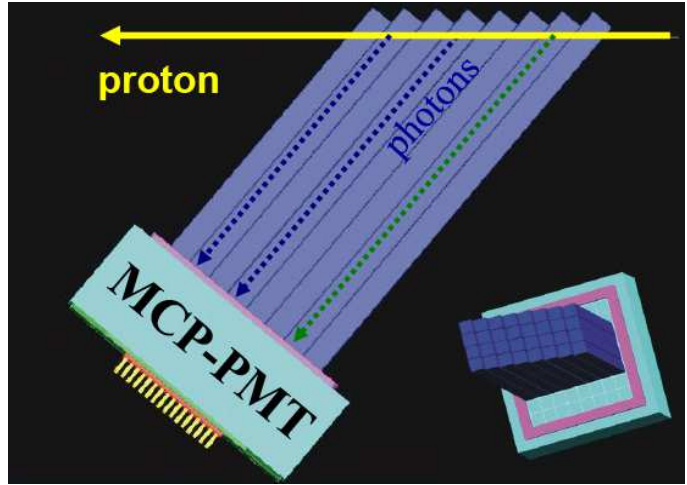}
\vspace*{-0.1in}
  \caption{A schematic diagram of the QUARTIC fast timing detector.
   } \label{fig5_timesys}
\end{figure}

\section{Timing detectors}

\subsection{Requirements and present achievement}

\begin{figure}
\centering
\includegraphics[width=3.5in]{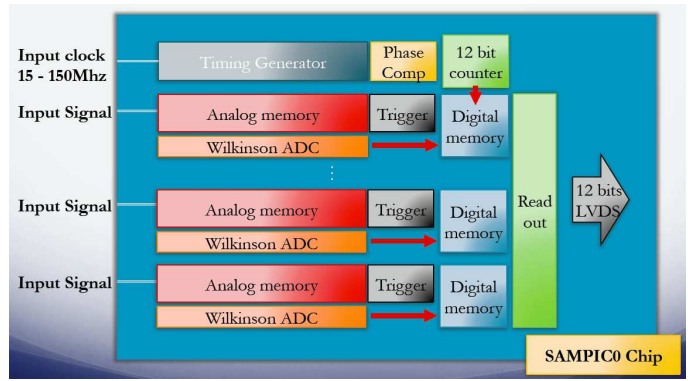}
\caption{Principle of the sampic chip.}\label{Figb}
\end{figure}

A fast timing system that can precisely measure the time difference between 
the two outgoing scattered protons is a key component of the AFP and 
CT-PPS detectors.  The 
time difference is equivalent to a constraint on the event vertex, thus 
the AFP and CT-PPS timing detectors can be used to reject overlapping 
background events by 
establishing that the two scattered protons did not originate from the same 
vertex that triggered the central ATLAS or CMS detectors. The final timing 
system should have the following characteristics~\cite{timing}: 
\begin{itemize}
\item 10 ps or better resolution (which leads to a factor 40 rejection on pile up 
background)
\item Efficiency close to 100\% over the full detector coverage
\item High rate capability (there is a bunch crossing every 25 ns at the nominal LHC)
\item Enough segmentation for multi-proton timing
\item High level trigger capability
\end{itemize}

\begin{figure}
\centering
\includegraphics[width=6.in]{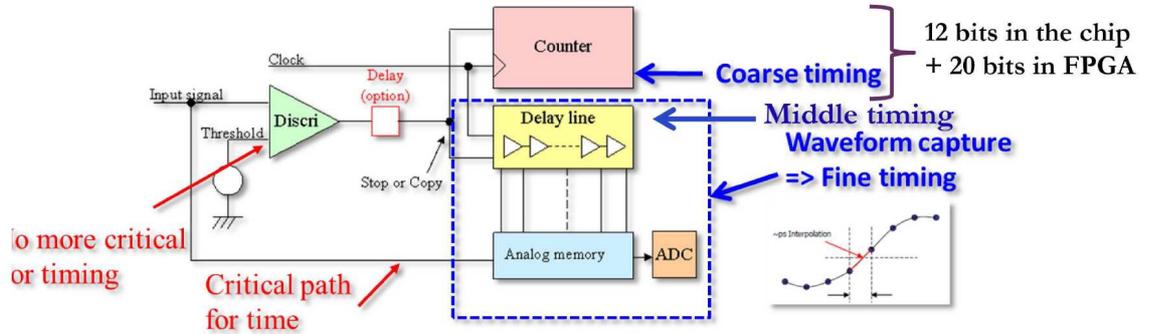}
\caption{Scheme of the sampic chip.}\label{scheme}
\end{figure}

Fig.~\ref{fig5_timesys} shows a schematic overview of the first proposed 
timing system in AFP, consisting of a quartz-based Cerenkov detector coupled to 
a microchannel plate photomultiplier tube (MCP-PMT), followed by the 
electronic elements  that amplify, measure, and record the time of the event 
along with a stabilized reference clock signal. The
QUARTIC detector consists of  an array of 8$\times$4 fused
silica bars ranging in length from about 8 to 12 cm and oriented at the
average Cerenkov angle.  A proton that is sufficiently deflected from the beam axis will pass 
through a row of eight bars emitting Cerenkov photons providing an overall time resolution that is 
approximately $\sqrt{8}$ times smaller than the single bar resolution of about 30 ps, 
thus approaching the 10 ps resolution goal. Prototype tests have generally been performed on one row 
(8 channels) of 5 mm $\times$ 5 mm pixels, while the initial detector is foreseen to have four rows
to obtain full acceptance out to 20  mm from the beam. The beam tests lead to a time resolution per
bar of the order of 34 ps. 
The upgraded design of the timing detector has equal rate pixels, 
and we plan to reduce the width of detector bins
close to the beam, where the proton density is highest. 

The CT-PPS also has a detector based on Cerenkov technology as the baseline 
proposal.  It has chosen  a  Cherenkov   L-bar Quartic design (Quartz 
Timing Cherenkov) with 5 $\times$ 4 equal to twenty 3 $\times$ 3 mm$^2$ 
independent channels.  They are read-out by SiPM photodetectors, 
relatively far from the beam, in a region where the neutron flux is $\sim$2
10$^{12}$ neq/cm$^2$ per 100 fb$^{-1}$. SiPM devices that tolerate this
radiation level are available, and already in use in the CMS
detector~\cite{cms3}. The SiPMs will probably require replacement after 100
fb$^{-1}$, which is feasible given the small number of devides involved. Two
such Quartic detectors fit inside a cylindrical roman pot, providing a combined
resolution of the order of 20 ps.

\subsection{Future developments}

\begin{figure}
\centering
\includegraphics[width=3.in]{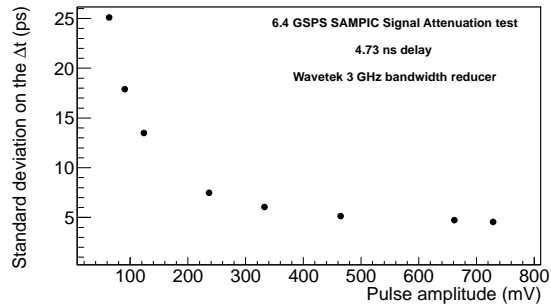}
\caption{Resolution of the SAMPIC chip. The resolution on the time difference between two
electronic channels was measured to be about 5 ps for a signal amplitude of 500
mV or higher, which means about 3 ps
resolution per channel.}\label{resol}
\end{figure}

\begin{figure}
\centering
\includegraphics[width=4.5in]{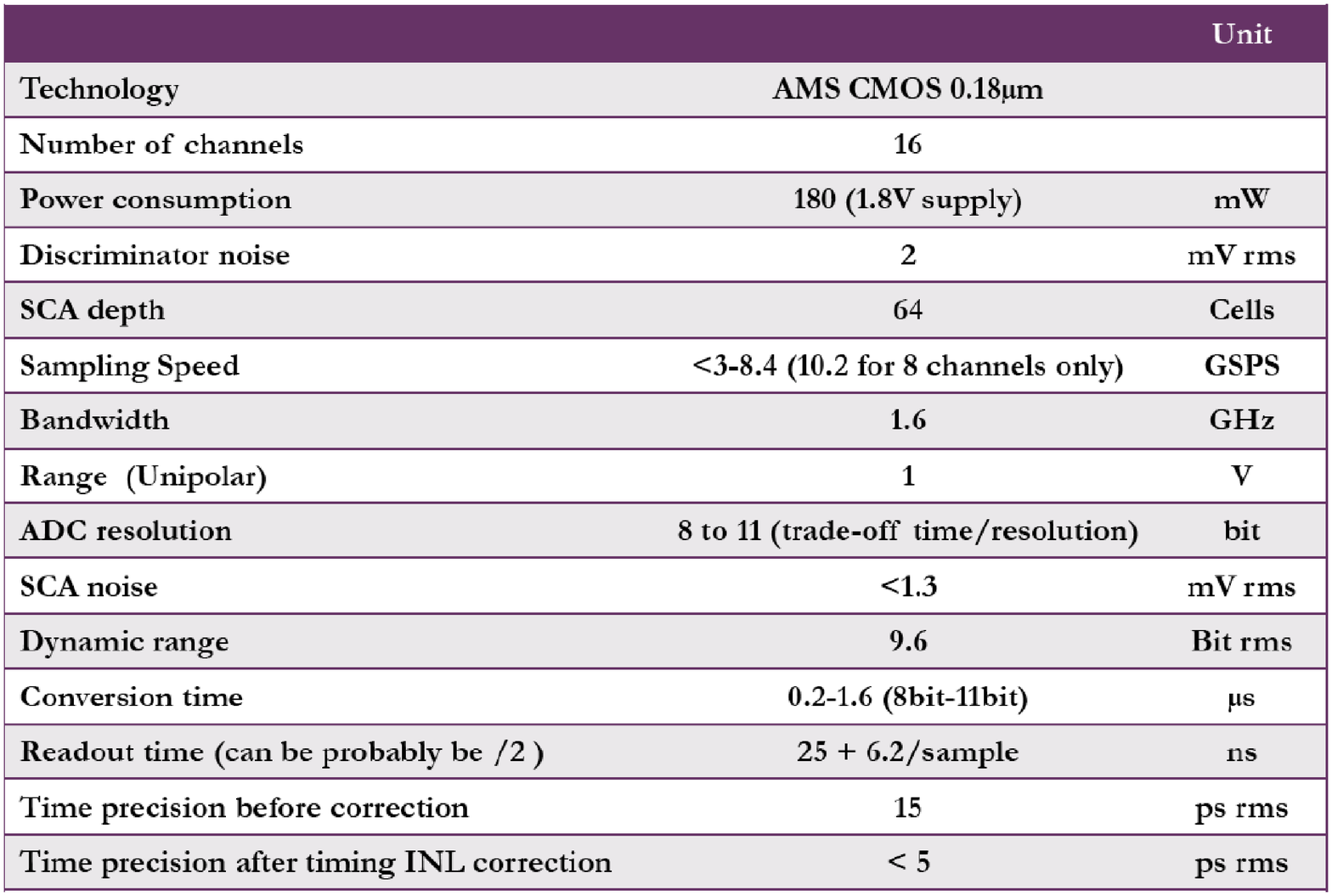}
\caption{Parameters of the SAMPIC chip.}\label{param}
\end{figure}

At higher luminosity of the LHC (phase I starting in 2019), higher pixelisation of the timing detector
will be required in order to fight against high pile up 
environment~\cite{chris}. For this sake, a R\&D phase to develop timing 
detector developments based on Silicon sensors~\cite{nicolo}, and
diamonds~\cite{gabriele} has started. This new R\&D aims at installing a prototype of such detector
at the LHC in the TOTEM experiment as soon as they are available.
In parallel to this sensor R\&D, a new timing readout chip has
been developed in Saclay. It uses waveform sampling to reach the best possible 
timing resolution: single-threshold and multi-threshold circuits are 
much more affected by the negative effects of time walk and jitter on the 
time resolution.. The
aim of this chip called SAMPIC~\cite{sampic}  (see Fig.~\ref{Figb}
and Fig.~\ref{scheme} for a schematic view of SAMPIC) is to 
obtain sub 10 ps timing resolution, 1GHz input bandwidth, no dead
time at the LHC, and data taking at up to 10.2 Gigasamples per second. 
The waveform TDC is a new concept that can reconstruct a signal by very fast
sampling. Inside SAMPIC, the timing measurement is performed in three step
hierarchy:  
(i) the coarse one using
a Timestamp Gray Counter (6 ns step), (ii) followed by a medium one when the DLL is locked on the
clock to define the region of interest (150 ps step) and  (iii) the fine one where
sampling is done in the region of interest (few ps resolution). 
To test the ultimate resolution of SAMPIC in ideal condition, an extensive 
testing campaign has been carried out on the chip itself.
Fig.~\ref{resol}
describes the results on the time difference resolution difference as a function
of the signal amplitude
obtained splitting one signal into two, and delaying one by
4.73 ns.
After calibration, we
obtain a resolution on the time difference better than 5 ps, which means a resolution per
channel of about 3 ps for a signal amplitude larger than about 500 mV. 
The parameters of the SAMPIC chip are given in
Fig.~\ref{param}.

\begin{figure}[htb]
\centerline{%
\includegraphics[width=8.5cm]{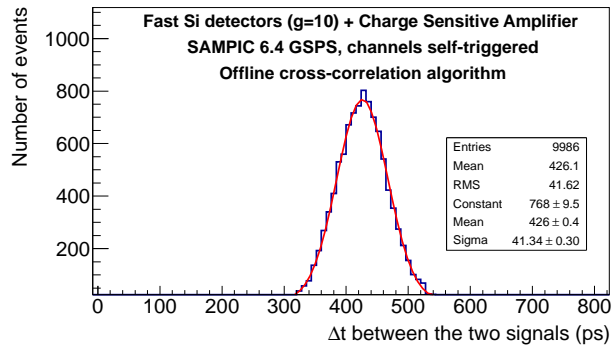}}
\caption{Preliminary results on timing resolution using SAMPIC and a Si detector
using a laser.}
\label{Fig5}
\end{figure}

SAMPIC has also been tested in combination with a new type of silicon 
detectors, the so called Ultra-Fast Silicon Detectors~\cite{cms4}, that employ 
internal multiplication to achieve a larger signal, well suited for 
timing applications. Employing a split 1064 nm laser beam, tuned to reproduce 
the signal amplitude of a minimum ionizing particle, we illuminated pairs of 
sensors and computed, using SAMPIC, the jitter of each sensor. Fig.~\ref{Fig5}
shows the result of this test: 
a promising resolution of 30 ps per channel. The parameters of the SAMPIC chip
are given in Fig.~\ref{param}.

Beam tests are
foreseen towards the Fall 2014 to further test SAMPIC together with a diamond and Si
detector. Further improvement of the SAMPIC chip will include the dead time
reduction using the ping-pong method.

The cost per channel is estimated to be
of the order for \$10 which a considerable improvement to the present cost of a
few \$1000 per channel,
allowing us to use this chip in medical applications such as PET imaging 
detectors. The holy grail of
imaging 10 picosecond PET detector seems now to be feasible: with a resolution better than 20 ps,
image reconstruction is no longer necessary and real-time image formation becomes
possible.

The PPS project has now been endorsed by the CMS and TOTEM collaboration at
least for the first phase at low luminosity. If everythings works as expected,
and the beam induced background (not easily simulated) is not found to be an
issue at 14 TeV, the project will be naturally approved to work at higher luminosity.
The AFP project is almost at the same stage, pending the approval at low
luminosities until enough resources are found within and outside the ATLAS
collaboration.

\section{Timimg detector optimisation for pile up rejection}  

In this section, we discuss possible optimisation of the timing detector in
terms of spacial resolution in order to reject pile up background. 

\subsection{Proton detection in the forward region}
The main source of background in the timing detectors is due to pile up events.
Intact protons may obviously originate from the diffractive and photon-exchange
events but also from additional soft interactions (pile up). For instance a
non-diffractive $WW$ event can be superimposed with two single diffractive soft
events with intact porons and it is important to be able to distinguish this
background from the event where both protons originate from the $WW$ vertex.
In order to
suppress this background, it is useful to measure precisely the proton
time-of-flights in order to know if the protons originate from the main event
hard vertex or not. 

Two parameters to build a detector are important to reject pile up:
\begin{itemize}
\item the precision of the proton time-of-flight, which is the timing detector
resolution. Typically a measurement of 10 ps gives a precision of 2.1 mm on the
vertex position
\item the pixelisation of the timing detector: at highest luminosity, the number
of intact protons per bunch crossing is high and in order to compute the
time-of-flight of each proton it is needed to have enough pixelisation or space
resolution so that each proton can be detected in different cells of the timing
detector. If two protons with different time-of-flight fall in the same cell,
the information is lost.
\end{itemize}

\subsection{Pixelisation of the timing detector}
In order to study the required pixelisation of the timing detector, we simulated
10 million minimum bias events (non diffractive, single diffractive and double
diffractive events) using the PYTHIA generator. The protons were
transported through the LHC magnets up to the proton detectors. Events are
characterised as no tagged (NT), single tagged (ST) and double tagged (DT)
depending on the number of protons in the forward proton detector acceptance.
For one minimim bias event, we get a probability of 97\% NT, $p=$1.6\% ST, and 
$q=$0.01\% DT. The multinomial distribution was adopted to simulate pile up
since we assume that the different interactions are independant~\cite{matthias}. 
For a given number of pile up proton $N$, the probability
to have $N_L$ ($N_R$) protons tagged in the left (right) side only, $N_B$ protons on
both sides and $N_N$ protons not tagged reads:
\begin{eqnarray}
P(N_B,N_L,N_R,N_N) = \frac{N!}{N_B!N_L!N_R!N_N!} p^{N_L} q^{N_B} p^{N_R}
(1-2p-q)^{N_N}
\end{eqnarray}
and the probability of no proton tagged, of at least one proton tagged 
on the left side, and of at
least one proton tagged on both sides reads
\begin{eqnarray}
P_{no~hit} &=& P(0,0,0,N) = (1-2p-q)^N \\
P_{hit~left} &=& \sum_{N_L=1}^{N} P(0,N_L,0,N-N_L) = (1-p-q)^N - (1-2p-q)^N \\
P_{double~hit} &=&1-P_{no~hit}-2P_{hit~left}=1+(1-2p-q)^N-2(1-p-q)^N
\end{eqnarray}

\begin{figure}[htb]
\centerline{%
\includegraphics[width=6.5cm]{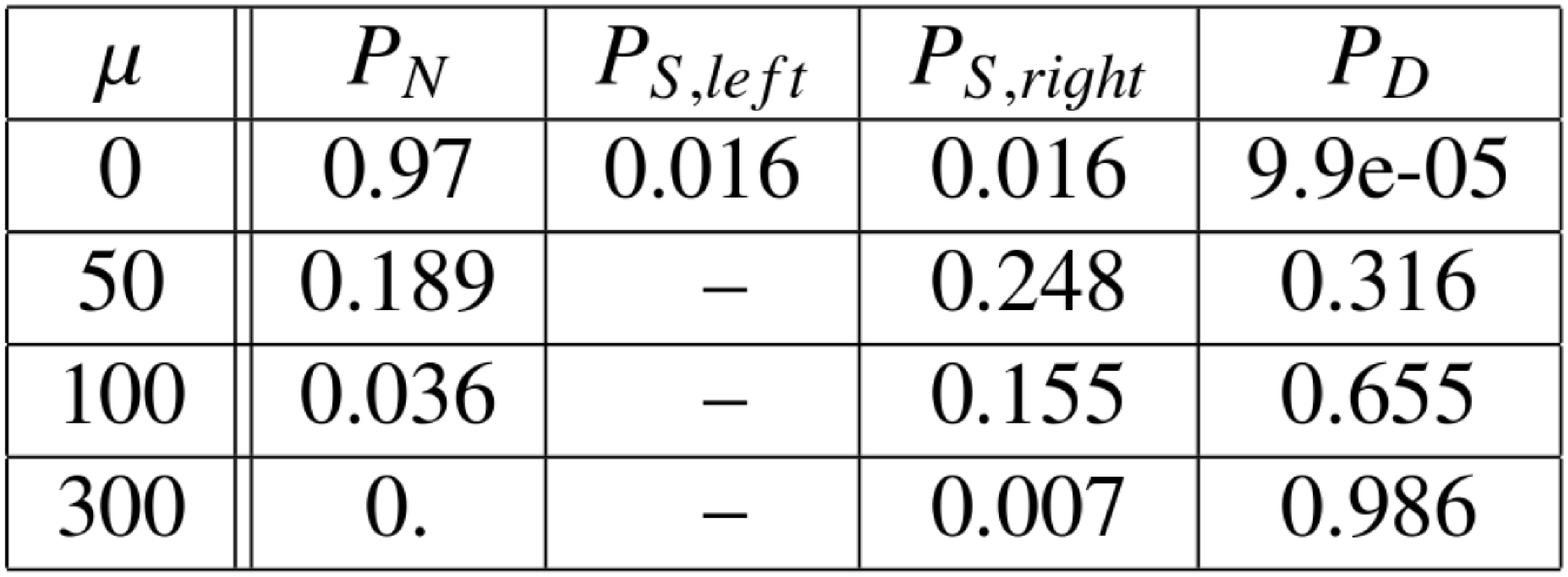}}
\caption{No tag, single tag on the left or right side (same by definition) and
double tagged probablity.}
\label{Fig2}
\end{figure}

\begin{figure}[htb]
\centerline{%
\includegraphics[width=10.5cm]{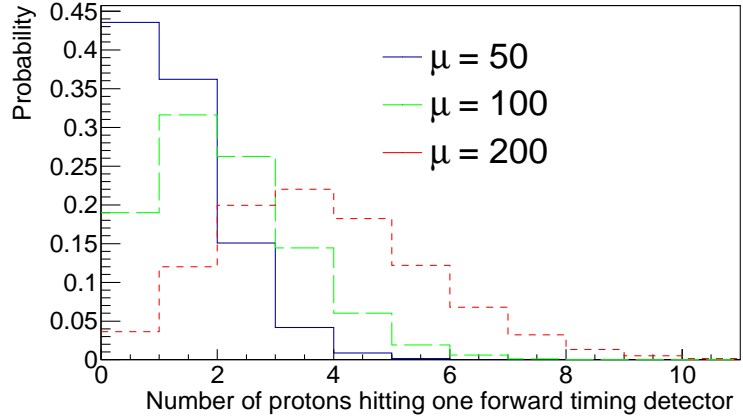}}
\caption{Probability of getting 0, 1, 2, 3... intact protons on one side of the
detector for 3 different values of $\mu$.}
\label{multi}
\end{figure}

\begin{figure}[htb]
\centerline{%
\includegraphics[width=10.5cm]{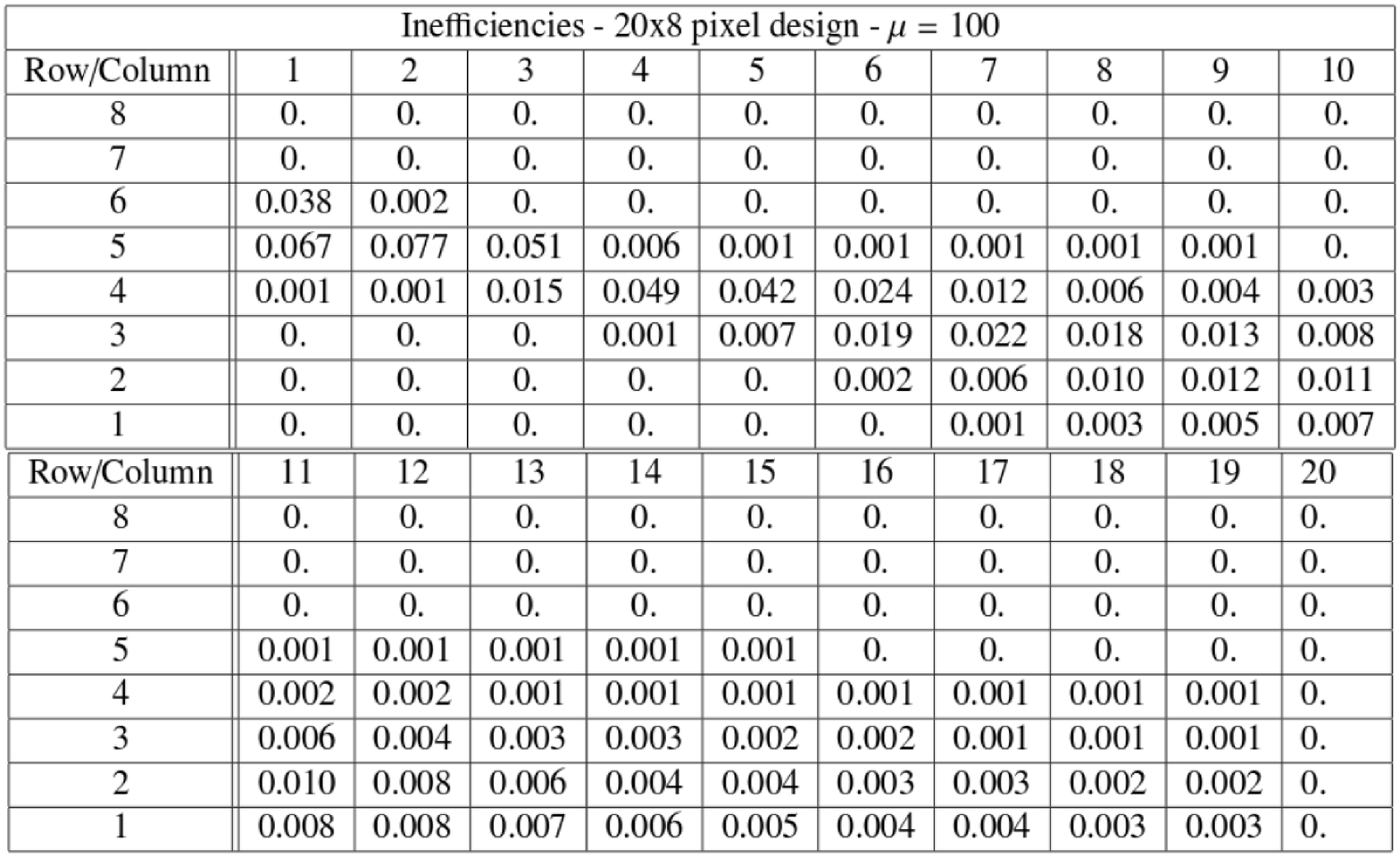}}
\caption{Probability of more than 1 proton to fall in a given pixel of the
timing detector.}
\label{Fig3}
\end{figure}

The hit probabilities can then be calculated for various pile up values (see
Fig.~\ref{Fig2}). For a pile up $\mu=$50 (100) for instance, the probability of
no tag is 19\% (3.6\%). Let us note that this simplified approach does not work
at very high pile up (300 for instance) since we neglected the cases when two
or more protons from pile up events can hit one side of the detector at
the same time. In order to illustrate this, the percentage of events corresponding to 0, 1, 2, 3... protons
on one side for $\mu=$50, 100 and 300 is given in Fig.~\ref{multi}.
This leads to larger inefficiencies that can be taken account in
a more refined approach or by a full pile up simulation, which was performed for
the $\gamma \gamma \gamma \gamma$ quartic anomalous coupling study.
The detector needed to detect intact protons has a
coverage of about 2 cm$\times$2 cm and is located 15$\sigma$ from the beam. 
The inefficiency of such a detector assuming
20$\times$8 pixels is given in Fig.~\ref{Fig3}. The numbers displayed in
the table correspond to the probability of getting one proton or more in a given
pixel for $\mu=$100 or a 20$\times$8 pixellised detector. The upper limit on the
inefficiencies if the order of 8\% for the pixels closest to the bins, but is
found negligible for pixels further away which measure higher mass diffractive
objects. For comparison, the inefficiences for $\mu=$50 is about half, and
vertical bar detectors lead to larger inefficiencies between 10 and 20\% on a
large part of the detector (a 7 bar detector with 2 mm width for the first bar
and 3.25 mm for other bars, leads to inefficiences betweem 8\% and 19\% for the
first 6 bars). It is also worth mentioning that this study only includes physics
backgrounds and not beam-induced backgrounds which are not in the
simulation. Recent results from TOTEM show that the beam-induced backgrounds
have the tendency to be high and located in the pixels closest to the
beam~\cite{totem}, and this is why a full pixelised detector is preferable to
bars.

\section*{Acknowledgments}
The study of the timing detector optimisation was performed in collaboration
with Matthias Saimpert, Old\v{r}ich Kepka and Radek Zleb\v{c}ik.


\end{document}